\documentclass[iop, twocolumn, tighten]{emulateapj}

\newcommand{\numstar}{3294 }
\newcommand{\Teff}{T_\star}
\newcommand{\Rp}{R_\mathrm{p}}

\newcommand{\dsp}{a}

\newcommand{\Rs}{R_\mathrm{\star}}
\newcommand{\Trm}{\mathcal{T}}
\newcommand{\TP}{r_\mathrm{T}}
\newcommand{\QE}{r_\mathrm{Q}}
\newcommand{\FS}{r_\mathrm{PSF}}

\newcommand{\biom}{\mathrm{abs}}

\newcommand{\IA}{f_\mathrm{A}}
\newcommand{\Ip}{f_\mathrm{p}}
\newcommand{\Ileak}{f_\mathrm{L}}
\newcommand{\Isky}{f_\mathrm{sky}}

\newcommand{\Ssky}{S_\mathrm{sky}}

\newcommand{\IB}{f_\mathrm{B}}

\newcommand{\Is}{f_\mathrm{\star}}

\newcommand{\Np}{\mathcal{N}_\mathrm{p}}

\newcommand{\Nsky}{\mathcal{N}_\mathrm{sky}}
\newcommand{\NNp}{\mathcal{N}_\mathrm{p}}
\newcommand{\NNleak}{\mathcal{N}_\mathrm{L}}
\newcommand{\NNsky}{\mathcal{N}_\mathrm{sky}}
\newcommand{\Ns}{\mathcal{N}_\mathrm{\star}}

\newcommand{\Nabs}{\mathcal{N}_\mathrm{abs}}

\newcommand{\XY}{{\bf X}}
\newcommand{\XYP}{{\bf X^\prime}}

\newcommand{\Cont}{C}

\newcommand{\Woxy}{W_{\mathrm{1.27 \mu m}}}

\newcommand{\Texp}{T_\mathrm{exp}}
\newcommand{\DA}{D}

\newcommand{\kfactor}{K}
\newcommand{\EW}{W_\mathrm{line}}

\slugcomment{Accepted for publication in The Astrophysical Journal}
\shortauthors{Kawahara et al.}
\shorttitle{Biomarker in Exoplanets from Ground-Based Telescopes}

\begin{document}
\title{Can Ground-based Telescopes Detect The Oxygen 1.27 Micron Absorption Feature as a Biomarker in Exoplanets ?}

\author{Hajime Kawahara\altaffilmark{1},Taro Matsuo\altaffilmark{2}, Michihiro Takami \altaffilmark{3},  Yuka Fujii\altaffilmark{4}, Takayuki Kotani\altaffilmark{5}, Naoshi Murakami\altaffilmark{6}, Motohide Tamura \altaffilmark{5}, and Olivier Guyon\altaffilmark{7,8}} 
\altaffiltext{1}{Department of Physics, Tokyo Metropolitan University,
  Hachioji, Tokyo 192-0397, Japan}
\altaffiltext{2}{Kyoto University}
\altaffiltext{3}{Institute of Astronomy and Astrophysics, Academia Sinica. P.O. Box 23-141, Taipei 10617, Taiwan, R.O.C.}
\altaffiltext{4}{Department of Physics, The University of Tokyo, Tokyo 113-0033, Japan}
\altaffiltext{5}{National Astronomical Observatory of Japan, 2-21-1 Osawa, Mitaka, Tokyo 181-8588, Japan}
\altaffiltext{6}{Division of Applied Physics, Faculty of Engineering, Hokkaido University, Sapporo, Hokkaido, 060-8628, Japan}
\altaffiltext{7}{National Astronomical Observatory of Japan, Subaru Telescope, Hilo, HI 96720, USA}
\altaffiltext{8}{Center for astronomical Adaptive Optics, Steward Observatory, University of Arizona, 933 N Cherry Ave, Tucson AZ 85721}
\email{kawa\_h@tmu.ac.jp}

\begin{abstract}
The oxygen absorption line imprinted in the scattered light from the Earth-like planets has been considered the most promising metabolic biomarker of the exo-life. We examine the feasibility of the detection of the 1.27 $\mu$m oxygen band from habitable exoplanets, in particular, around late-type stars observed with a future instrument on a 30 m class ground-based telescope. We analyzed the night airglow around 1.27 $\mu$m with IRCS/echelle spectrometer on Subaru and found that the strong telluric emission from atmospheric oxygen molecules declines by an order of magnitude by midnight. By compiling nearby star catalogs combined with the sky background model, we estimate the detectability of the oxygen absorption band from an Earth twin, if it exists, around nearby stars. We find that the most dominant source of photon noise for the oxygen 1.27 $\mu$m band detection comes from the night airglow if the contribution of the stellar PSF halo is suppressed enough to detect the planet. We conclude that the future detectors for which the detection contrast is limited by photon noise can detect the oxygen 1.27 $\mu$m absorption band of the Earth twins for $\sim$ 50 candidates of the late type star. This paper demonstrates the importance of deploying small inner working angle efficient coronagraph and extreme adaptive optics on extremely large telescopes, and clearly shows that doing so will enable study of potentially habitable planets.
\end{abstract}

\section{Introduction}
 Oxygenic photosynthesis is currently the predominant source of energy for life on Earth:
\begin{eqnarray}
\mathrm{6 CO_2 + 12 H_2 O \to C_6 H_{12} O_6 + 6 H_2 O + 6 O_2}.
\end{eqnarray}
To be more specific, oxygen molecules are generated by the transfer of electrons from water: 
\begin{eqnarray}
\mathrm{2 H_2 O} + 4 n \, \gamma \to \mathrm{O_2 + 4 H^+ + 4 e^-},
\end{eqnarray}
where $\gamma$ indicates a photon and $n$ specifies the $n$-photon process of the photosynthesis. Terrestrial oxygenic photosynthesis has the $n=2$ photon process to utilize the main band frequency of the sunlight $\lambda = 0.4- 0.7 \mu$m (Photosynthetically Active Radiation; PAR). The abiotic production of oxygen molecules is difficult since a main source of abiotic oxygen molecules on the Earth, the direct photolysis of water molecules, is only effective for energetic photons in UV, $\lambda < 180$ nm \citep{2011AsBio..11..335L}. Hence oxygen and ozone absorptions have been considered as the most promising biosignatures for the search for life on terrestrial exoplanets \citep[ and references therein]{1980ASSL...83..177O,1986Natur.322..341A,1993A&A...277..309L,2002A&A...388..985S,kaltenegger2010,2011AsBio..11..335L}.

Space-based coronagraph missions proposed so far have been aimed at detecting the oxygen 0.76 $\mu$m and 0.69 $\mu$m bands from habitable planets mainly around G-type stars \citep[e.g. TPF-C][]{2009arXiv0911.3200L}. Since it is difficult for space missions to utilize large telescope (a $\le$ 4 m aperture is typically assumed), direct imaging of habitable planets around late-type stars is more challenging due to their small angular separation from the host star. On the other hand, 30 m class or larger ground-based telescopes have also been under consideration for next generation missions including the Thirty Meter Telescope (TMT), the European Extremely Large Telescope (E-ELT) and Giant Magellan Telescope (GMT). Several detectors for direct imaging of exoplanets have been proposed for these telescopes \citep[e.g.][]{2006SPIE.6272E..20M,2010SPIE.7735E..81K}. The advantages and disadvantages of the direct imaging of the ground-based telescope have been widely discussed so far \citep[e.g.][]{2009astro2010T..12M,2009arXiv0910.4339A,2010lyot.confE..46H}. While several authors pointed out that the Extreme Adaptive Optics (ExAO) has difficulty to achieve the contrast $10^{-8}$, which is not sufficient for the Earth-like planet detection \citep[e.g.][]{2006IAUS..232..149S}, several authors claimed that the postprocessing with ExAO can overcome the $10^{-8}$ limit and achieve the contrast below $10^{-9}$. Moreover, the contrast of a habitable Earth-size planet around late-type star is much more favorable than that of G-type star ($10^{-10}$), for instance, $\sim 10^{-7} - 10^{-9}$ for late-type stars since the reflected-light luminosity is more or less constant among all habitable planets due to the temperature constraint, while the luminosity of stars differs by several orders of magnitude between spectral types.

For ground-based direct imaging, near-infrared (NIR) observation is preferable over visible thanks to better ExAO performance. Besides absorptions at visible bands, oxygen molecules also have the absorption feature in the NIR band, at 1.27 $\mu$m with a 0.02 $\mu$m width, which is the second strongest absorption feature next to 0.76 $\mu$m band. While the detectability of this band by ground-based telescope has been studied in the context of the transmission spectroscopy of exoplanets around late-type stars \citep{2011ApJ...728...19P}, it has not been considered in the context of the reflected light in detail.

Recently several missions to aim to detect the habitable planets around late-type stars have been proposed \citep[e.g.][]{2010SPIE.7735E.264M, 2010SPIE.7735E..81K}. The key concept to detect habitable planets from the ground is the combination of a coronagraph and subsequent post processing \citep[e.g.][]{2011PASP..123...74H}. Although the key technology of ExAO, coronagraphs and post-processing have not yet demonstrated the performance level required to detect habitable planets, many proposed techniques are currently under development aiming toward reaching the photon noise limit \citep[][references therein]{2006MNRAS.373..747P, 2006SPIE.6272E..72S,2010A&A...509A..31G,2010aoel.confE9005C,2010aoel.confE5008B,2011PASP..123.1434V}. For instance, \citet{2010aoel.confE5008B} presented simulation results of the self coherent camera (SCC) assuming E-ELT and showed that the SCC can achieve the photon noise limited when the magnitude is below $\sim 7$ mag for the H band. In addition, speckle nulling technique,  which is not a post-processing method, is also a promising method to detect a habitable planet around a late-type star. Hence it is crucial for astrobiology on exoplanets to know what kind of science at the habitable planet is possible by ground-based telescopes assuming that  an idealistic photon-limited detection is reality in the near future. 

In this paper, we focus on the 1.27 $\mu$m oxygen absorption band imprinted in reflection lights as a biomarker of oxygenic photosynthesis and examine its detectability with the above 30-40 m class telescopes (Extreme Large Telescopes; ELTs) with near-future idealistic instruments against the speckle noise.

The rest of the paper is organized as follows. We first describe the outline of the absorption detection and clarify the main source of noise in \S 2. In particular we estimate the intensity of the sky background by analyzing the real data of the sky on Mauna Kea. With several nearby star catalogs, we estimate the feasibility of the oxygen 1.27 $\mu$m detection assuming an Earth-twin at the inner edge of habitable zone (IHZ) in \S 3. In \S 4, we discuss the availability of the stellar radiation to use the oxygenic photosynthesis for our sample. Finally we summarize our results in \S 5.

\section{Direct Imaging of Earth-like Planets by Ground-based Telescopes}

\begin{figure}[!tbh]
\includegraphics[width= \linewidth]{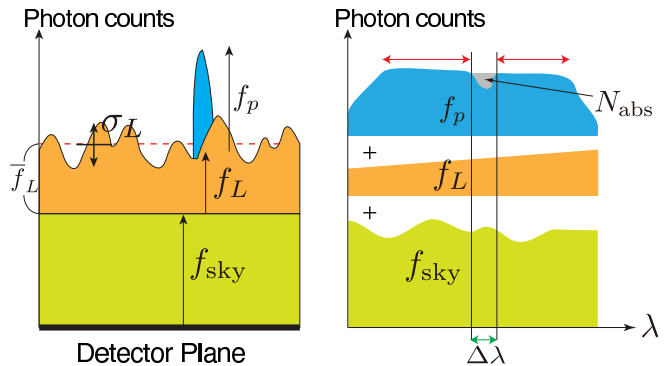}
  \caption{Schematic view of the planet detection (left) and the absorption band detection (right). The photons from the planet, the leakage of the main star, and the sky background (airglow) are painted by blue, orange, and green, corresponding to their flux $\Ip$, $\Ileak$, and $\Isky$. Green and red arrows indicate the range for the absorption bands and for the continuum determination. \label{fig:scheme}}
\end{figure}

The detectability of the planet itself and the detection of absorption lines are different. The former is in many cases dominated by the speckle noise from the leakage of the main star, as shown in the left panel of Figure \ref{fig:scheme} (we use the term "leakage" to describe the contribution of the stellar PSF halo throughout this paper). Thus, the signal-to-noise ratio is $\Ip/\sigma_\mathrm{L}$, where $\Ip$ is the flux of the planet and $\sigma_\mathrm{L}$ is the standard deviation of the leakage $\Ip$. The sky background is expected to be uniform on the detector plane, and its non-uniformity and photon noise are negligible compared with the speckle noise. On the other hand, once the planet has been detected, the photon noise of the skybackground $\Isky$ , the planet signal, and the leakage is the main source of the statistical error of the band detection (right panel of Figure \ref{fig:scheme}). In this section, we concentrate on the detectability on ground of the 1.27 $\mu$m band assuming that planets themselves are detectable. We will discuss the relation between the detectability of planets and the oxygen band with the nearby star catalogue in \S 3. 

\subsection{Signal-to-Noise Ratio of Detection of the Absorption Band}
Here we summarize our observational strategy to detect biomarker absorption bands.
The flux within the solid angle of the aperture $\Delta \Omega$ where the planet is located is expressed as 
\begin{eqnarray}
\IA(\lambda) = \Trm(\lambda) \Ip(\lambda)  + \Isky(\lambda) +  \Trm(\lambda) \Ileak(\lambda), 
\end{eqnarray}
where $\Trm (\lambda) $ indicates the atmospheric transmission, $\Ip(\lambda)$ is the reflection light from the planet,  $\Ileak(\lambda)$ is the leakage from the main star (speckle noise and halo), $\Isky(\lambda) = \Ssky \Delta \Omega$ is the sky background within the aperture due to the night airglow, and $\Ssky(\lambda)$ is the surface brightness of the sky background.

Simultaneous observation of the planet and the main star is essential to eliminate the effect of atmospheric transmission variation. This requirement is compatible with nulling coronagraph concepts \citep[e.g.][]{2004SPIE.5487.1296S,murakami}. The flux from the observation of the main star is expressed as 
\begin{eqnarray}
\IB(\lambda) = \Trm(\lambda) \Is(\lambda) + \Isky(\lambda) \approx  \Trm(\lambda) \Is(\lambda),
\end{eqnarray}
where $\Is (\lambda)$ is the stellar flux. Since $\Isky (\lambda)$ can be simultaneously obtained on the outer side of the detector plane, we can estimate the summation of reflectivity and the contrast difference
\begin{eqnarray}
\label{fig:jcont1}
J(\lambda) &\equiv& \frac{\IA (\lambda) - \Isky (\lambda) }{\IB (\lambda)}  \\
\label{fig:jcont}
&\approx& \frac{\Ip (\lambda)}{\Is (\lambda)} + \Cont_\XY (\lambda),  \\
\Cont_\XY (\lambda) &\equiv& \frac{\Ileak(\lambda)}{\Is (\lambda)}
\end{eqnarray} 
Here we assume $\Cont_\XY(\lambda)$ does not contain absorption-like feature. We discuss this possibility in \S \ref{ss:fps}.

Since the photon noise in $\IA (\lambda)$ dominates the statistical noise of $J(\lambda)$, we define the signal to noise ratio of the absorption band as
\begin{eqnarray}
\label{eq:abssn}
(S/N)_{\biom} \equiv \frac{\Nabs}{\sqrt{\NNsky + \NNp + \NNleak}},
\end{eqnarray}
where  $\NNsky$, $\NNp$, and $\NNleak$ are the photo-electrons of the sky background, the planet, and the leakage within the absorption band width $\Delta \lambda$  at the band center $\lambda_c$. The photo-electrons in the absorption band $\Nabs$ (Fig. [\ref{fig:scheme}] right) are
 related to the equivalent width of the band $\EW$ as
\begin{eqnarray}
\Nabs  &=& \Np \, \frac{\EW}{\Delta \lambda} \left(1 - \frac{\EW}{\Delta \lambda}\right)^{-1},
\label{eq:Nabs}
\end{eqnarray}
where $\EW$ is the equivalent width of the band.

Equation (\ref{eq:abssn}) implicitly assumes that the continuum level, as indicated by red arrows in Figure \ref{fig:scheme}, can be determined without uncertainty. While this is not the case for real observation, the wavelength width available to determine continuum (width of red arrows) is 10 times larger than the absorption band width (width of green arrow) for the 1.27 $\mu$m band (see the mock spectra in Bottom Panel of Figure \ref{fig:spectra}). Hence we ignore the uncertainty of continuum level. We note that an OH lines suppresser can reduce this uncertainty further. In Equation (\ref{eq:abssn}), we also ignore the statistical noises from the denominator of Equation (\ref{fig:jcont1}) due to the stellar flux $\IB(\lambda)$ and the offset observation $\Isky (\lambda)$ since $\IB(\lambda)$ is much larger photon counts and we can sum up the wider region on the detector than $\Delta \Omega$ for $\Isky(\lambda)$.

\subsection{Reflection from the Planet and Strength of the Oxygen 1.27 $\mu$m Band}
The reflected light from the planet under Lambert diffuser assumption is expressed as 
\begin{eqnarray}
\Ip (\lambda) &=& \frac{2 \phi(\beta)}{3} A(\lambda) \left(\frac{R_p}{\dsp}\right)^2 \Is (\lambda), \\
\phi(\beta) &\equiv& [\sin{\beta} +  (\pi - \beta) \cos{\beta}]/\pi, 
\end{eqnarray}
where $A(\lambda)$ is the Bond albedo, $R_p$ is the planetary radius, and $\dsp$ is the star-planet distance. The Larmbert phase function $\phi (\beta) $ is characterized by the phase angle $\beta = \angle \mathrm{(star-planet-observer)}$.  We estimate the photo-electron from the planet as
\begin{eqnarray}
\Np &\approx& \kfactor  \, \FS \Trm(\lambda_c)  \, \frac{\lambda_c}{h c} \Ip(\lambda_c)  \, \Delta \lambda, \\
\kfactor &\equiv& \pi \left( \frac{\DA}{2} \right)^2 \Texp \TP \QE,
\label{eq:pcplanet}
\end{eqnarray}
where $\DA$ indicates the effective diameter of the telescope, , $\Texp$ is the exposure time,  $\TP$ is the throughput of the optical system, $\QE$ is the quantum efficiency, and $h$ and $c$ are the Planck constant and the speed of light. The $\FS$ is the encircled energy, which quantifies what percentage the planet flux is within the analyzed region. We assume a circular region of radius $\lambda/\DA$, that is, $\Delta \Omega = \pi (\lambda/\DA)^2$). Since Fraunhofer diffraction gives 83\% photons in this range with Strehl ratio=1, we assume $\FS=0.7$ assuming Strehl ratio $\sim 0.9$.

The oxygen 1.27 $\mu$m band of the reflection from the Earth has been measured by Earthshine observation, $\Woxy = 2.42 \pm 0.30$ nm \citep{2009Natur.459..814P}. The 1.27 $\mu$m band observed in Earthshine consists of mixture of the main source  $\mathrm{O}_2$ lines and the minor sources from atmospheric dimers (weakly bounded complex of two molecules by Van del Waals force) of $\mathrm{O_2 \cdot O_2}$ and $\mathrm{O_2 \cdot N_2}$ \citep[e.g.][]{2009Natur.459..814P}.

We also estimated the equivalent width by creating mock spectra with a radiative transfer code. The planetary reflection spectra are, in reality, slightly shifted by the peculiar velocity of the planet, while the Earthshine spectra are not. To take this effect into account, we computed the high resolution spectra ($d \lambda=$0.01 nm) using the radiative transfer code,  libradtran \citep{mayer2005technical} with the line-by-line scheme (LBL). We used the line-by-line optical properties for the US-standard atmosphere generated by genln2 \citep{1992ggll.rept.....E} based on HITRAN96 \citep{rothman1999hitran}, which is provided by the libradtran website \footnote{http://www.libradtran.org}.  We divide the planetary surface to 192 equal area facets  by HEALPix \citep{2005ApJ...622..759G} and compute the contribution from each facet assuming a planet with $\beta = 90^\circ$. The stellar spectra template for the effective temperature 3500 K and the solar-metalicity is used for the incident flux, which is taken from \citet{2005A&A...443..735C}. We also assume that $d=$ 5 pc, 0.4 solar radius for the stellar radius, and the star-planet distance 0.15 AU, which corresponds to the IHZ. We summed the spectra for the clear sky and cloudy sky (optical depth is set to 15) under the ground albedo 0.1 so as to make the spectra with the cloud cover fraction 0.5, which yields the albedo of the planet $\sim 0.3$.

\begin{figure}[!tbh]
\includegraphics[width=\linewidth]{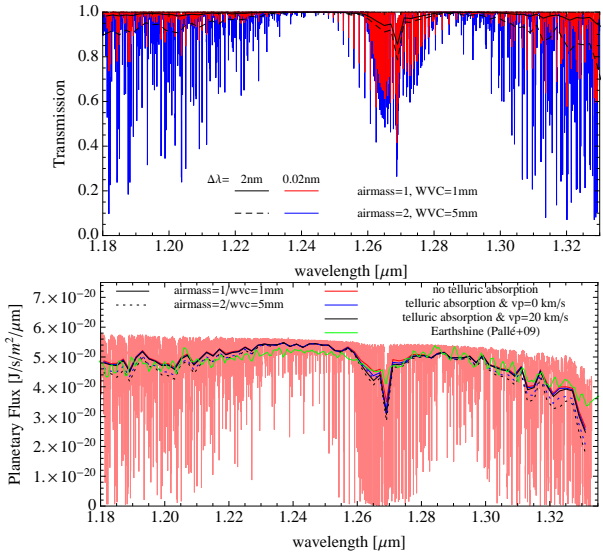}
  \caption{Top panel: Atmospheric transmission (airmass=1) at Mauna Kea taken from the Gemini website, which was generated by the ATRAN modelling software. Bottom panel: the mock spectra of the Earth-twin around an M-type star, created by solving radiative transfer with the Line-By-Line scheme. The flux with the spectral resolution 0.01 nm is shown by a thin line. Thick curves are binned in 2 nm interval, corresponding to flux without the telluric atmospheric absorption (red), with the absorption and zero peculiar velocity (blue), and with the absorption and 20 km/s peculiar velocity (black). Figure \ref{fig:at2} displays the fine structure of these three curves around 1.27 $\mu$m. \label{fig:spectra} For reference, we also plot the spectra taken from the Earthshine by green line \citep{2009Natur.459..814P}.}
\end{figure}

The planetary spectra are combined with the atmospheric transmission at the Mauna Kea assuming the peculiar velocity of the planet $v_p=0$ km/s and $v_p=20$ km/s. The latter corresponds the peculiar velocity of the Sun to the mean Galaxy rotation. We consider two extreme cases for the airmass and the water vapor column (WVC) of the telluric absorption $\Trm (\lambda) $, (airmass=1.0, WVC=1mm) and (airmass=2.0, WVC=5mm). The transmission templates at Mauna Kea site, as shown in the top panel of Figure \ref{fig:spectra}, were taken from the Gemini website \footnote{http://www.gemini.edu/sciops/telescopes-and-sites/observing-condition-constraints/ir-transmission-spectra}, which were generated by the ATRAN modelling software (Lord, S.D. 1992, NASA Technical Memor. 103957). 

\begin{figure}[!tbh]
\includegraphics[width= \linewidth]{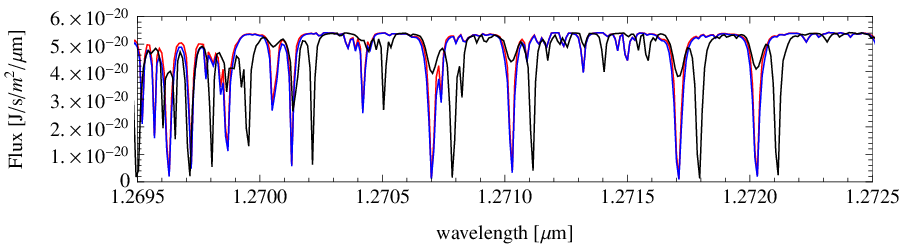}
\includegraphics[width= \linewidth]{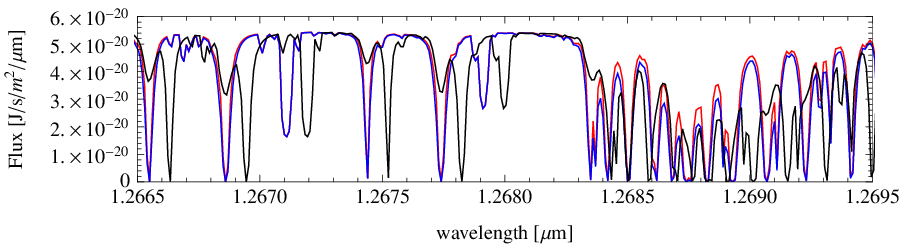}
  \caption{Fine structure of planetary mock spectra around 1.27 $\mu$m in Bottom Panel of Figure \ref{fig:spectra}. The red curve indicates the spectra without atmospheric absorption. The spectra with atmospheric absorption and peculiar velocity of the planet are shown by blue ($v_p=0$km/s) and black curves ($v_p=20$km/s), respectively. \label{fig:at2}}
\end{figure}

The bottom panel of Figure \ref{fig:spectra} shows the planetary spectra with high resolution ($\Delta \lambda=0.01$ thin curve) and the low resolution ($\Delta \lambda=2$ nm; red) obtained by binning the high resolution one.  Binned spectra transmitted in telluric atmosphere are also plotted for $v_p=0$ km/s (blue) and for $v_p=20$ km/s (black). The absorption  feature of the spectra with $v_p=20$ km/s is slightly stronger than that with $v_p=0$ km/s. Figure \ref{fig:at2} enlarges these three spectra around 1.27 $\mu$m. As shown in Figure \ref{fig:at2}, the peculiar velocity of 20 km/s is larger than the typical width of each thin line, which displaces the planetary lines from the telluric lines. This is the reason why the spectra with the peculiar velocity has slightly stronger absorption. However, this displacement of the lines does not make significant change to the equivalent width of the 1.27 $\mu$m band (2.5 nm for $v_p=0$km/s and 2.7 nm for $v_p=20$ km/s in the range 1.26-1.28 $\mu$m) as shown in the bottom panel of Figure \ref{fig:spectra} and both cases have the equivalent width $\sim 2-3$ nm, which is consistent with the result of the Earthshine observation \citep[$\Woxy = 2.4$ nm, green line][]{2009Natur.459..814P}. Hence we conservatively take $\Woxy = 2$ nm as the fiducial value of the equivalent width of the oxygen 1.27 $\mu$m band throughout this paper. Since the depth of the oxygen bands depends on the path of light, mainly driven by clouds, we only consider the depth seen in the current Earth and we do not consider the other cases.


Although photochemical reactions might alter the atmospheric composition for different stellar types \citep{2005AsBio...5..706S}, we ignore these effects since we consider the aerobic environment and the amount of oxygen molecules is unlikely to change in this situation.

\subsection{Night Airglow: Oxygen and OH Emission Lines \label{ss:airglow}}

The night airglow is described as 
\begin{eqnarray}
\Nsky \approx \kfactor  \, \frac{\lambda_c}{h c}  \langle \Ssky(\lambda) \rangle_b \, \Delta \Omega \, \Delta \lambda,
\label{eq:pcsky}
\end{eqnarray}
 where $\langle \Ssky(\lambda) \rangle_{b}$ is average of $\Ssky (\lambda)$ over the band $b$. There are numerous OH lines in the J and H bands, which dominates the night airglow. Unless using a suppresser of these lines such as OHS \citep{2001PASJ...53..355I}, the average in the J band is $ \langle \Ssky (\lambda) \rangle_{J} = \int_J d \lambda \Ssky/\Delta \lambda_J \sim 10^{-15} \, \mathrm{[J/s/m^2/\mu m/arcsec^2]} $ for the best condition, airmass=1 and water vapor=1.0mm on Mauna Kea \footnote{http://www.gemini.edu/sciops/telescopes-and-sites/observing-condition-constraints/ir-background-spectra}, where we adopt $\Delta \lambda_J = 0.17 $ $\mu$m for the band width . However, for the sky background around the 1.27 $\mu$m, one must consider the precise surface brightness around 1.27 $\mu$m with width 0.1-0.2 $\mu$m, which includes the atmospheric $\mathrm{O_2}$ emission in addition to the OH lines. Hence we analyze the sky observation using archival data by IRCS/echelle \citep{1998SPIE.3354..512T} on Subaru telescope in November 24, 2007 (Table \ref{tab:ircs}).

\begin{table*}[!tbh]
\begin{center}
\caption{IRCS Data for the sky background analysis in Nov 24, 2007. \label{tab:ircs}}
  \begin{tabular}{ccccc}
   \hline\hline
Local time (HST) &  ID & Object & Airmass & Exposure/frame \\
   \hline
19:12-1:34 & IRCA00189116 - 00189182 & SDSSJ01408-0839 & 1.1-2.1 & 900 s \\
3:11-5:34 & IRCA00189237 - 00189255 & PSSJ1057+4555 & 1.2-1.7 & 900 s \\
   \hline
\end{tabular}
\end{center}
\end{table*}

\begin{figure*}[!tbh]
\includegraphics[width= \linewidth]{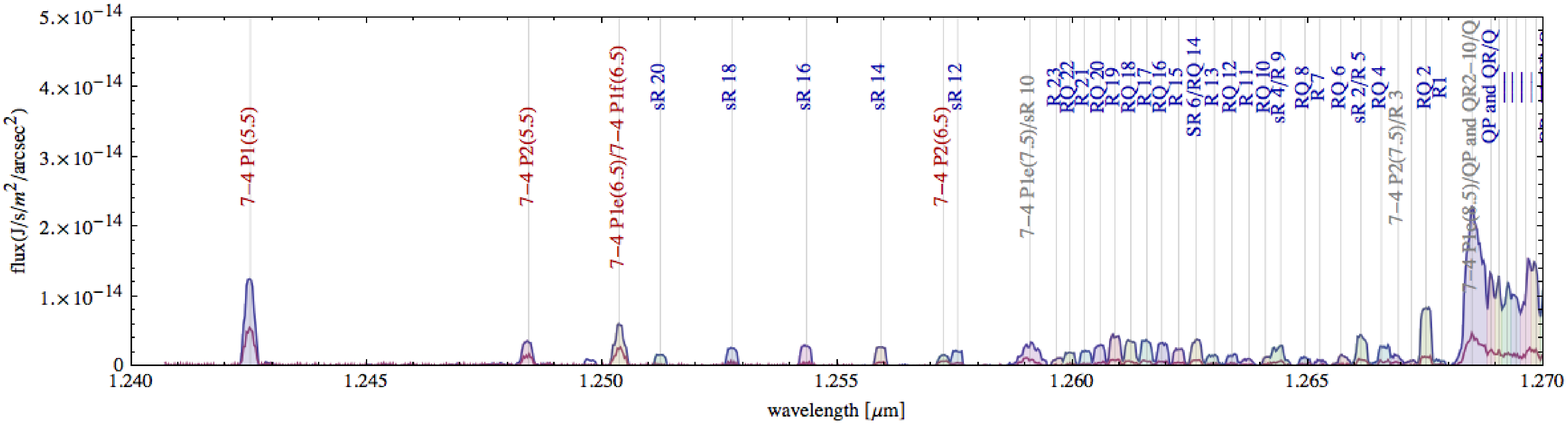}
\includegraphics[width= \linewidth]{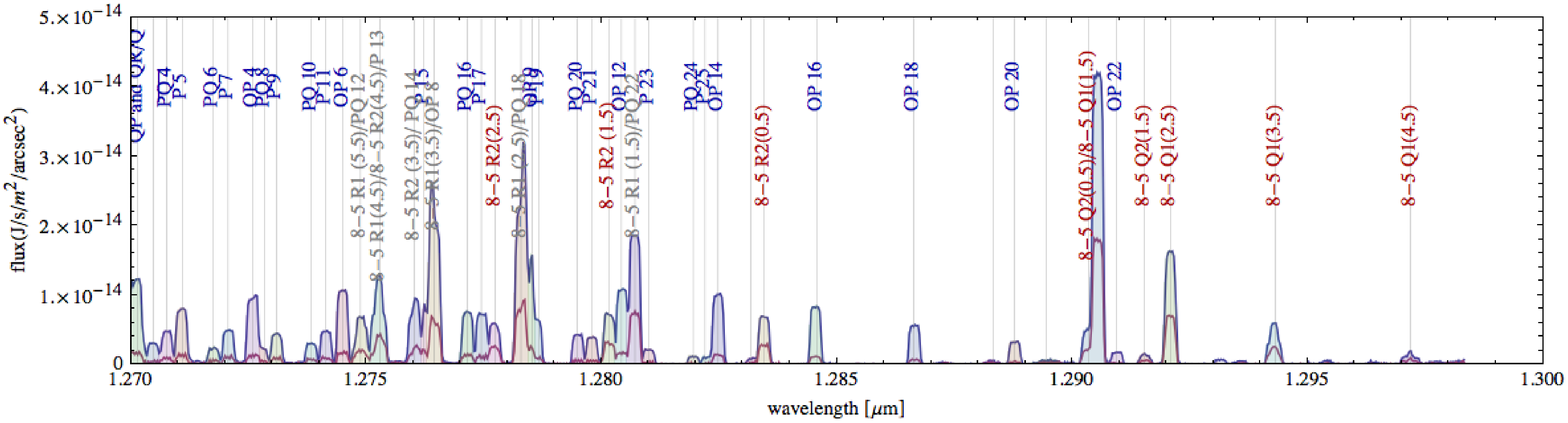}
  \caption{Sky background spectra at Mauna Kea obtained by IRCS/echelle on Subaru telescope. The blue and red curves indicate the spectra observed at 19:12 and 5:34. Thin vertical lines specify the line wavelength identified by a visual inspection. Blue, red, and gray labels beside the vertical lines were identified as the OH, $\mathrm{O_2}$, and mixture emission lines based on \citet{2000A&A...354.1134R}. \label{fig:spec}}
\end{figure*}

We extracted the spectra from 19:12 to 1:34  and from 3:11 to 5:34 (local time), while we could not use the data during 1:34-3:11 since each exposure time in this period is too short to detect OH/$\mathrm{O_2}$ lines. The spectra were obtained with a 0''.56 slit providing a spectral resolution of $\sim$5000. This is sufficient for resolving most of OH and $\mathrm{O_2}$ lines around 1.27 $\mu$m. The data were reduced via the following procedures. Each frame was dark subtracted and flat fielded; the median sky value in photo-electron (or the number of photons) along the slit was determined; and the spectrum was calibrated in flux (using spectra of standard stars) and in wavelength. Based on the OH/$\mathrm{O_2}$ line catalogue of \citet{2000A&A...354.1134R}, we identify lines in the 1.24-1.30 $\mu$m range by visual inspection and classify them by their origin, the OH line (labeled by red color), and the $\mathrm{O_2}$ line (blue color) and unresolved mixture of OH and $\mathrm{O_2}$ lines (gray) as shown in Figure \ref{fig:spec}. 

\begin{figure}[!tbh]
\includegraphics[width= \linewidth]{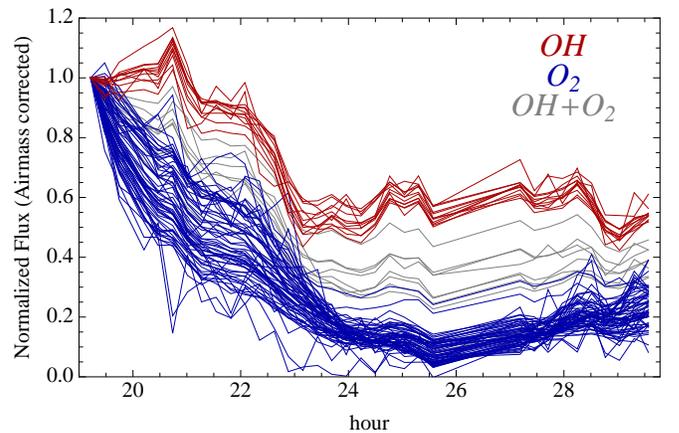}
  \caption{Nocturnal evolution of the sky emission lines normalized by value at 19:12 as a function of the local time (Hawaii Standard Time). The red and blue curves indicate OH and $\mathrm{O_2}$ lines, respectively. Gray curves are mixture of unresolved lines of the OH and $\mathrm{O_2}$ lines. The OH line shows a decrease by a factor of $\sim$ 2 in the OH emission throughout the night, while the $\mathrm{O_2}$ lines significantly decline by a order of magnitude.  \label{fig:lineev}}
\end{figure}

\begin{figure}[!tbh]
\includegraphics[width= \linewidth]{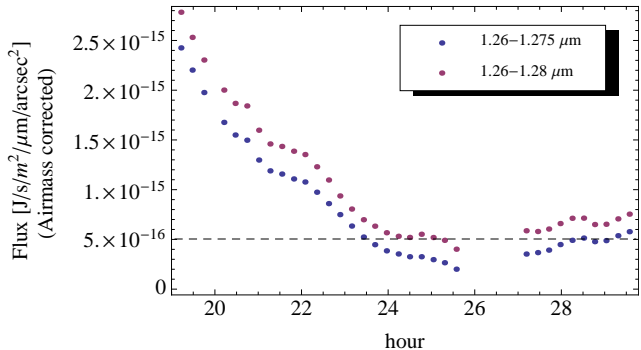}
  \caption{Surface brightness around the 1.27 $\mu$m band as a function of the local time. The red and blue points indicate the surface brightness of the sky background in November 24, 2007 of the 1.26-1.28 $\mu$m range and the 1.26-1.275 $\mu$m range, respectively. The latter range is selected so as to avoid strong OH lines present in the 1.275-1.280 $\mu$m range. The horizontal dashed line indicates our fiducial value. \label{fig:fluxsky}}
\end{figure}

Figure \ref{fig:lineev} is the nocturnal evolution of OH (red), $\mathrm{O_2}$ (blue), and mixture (gray) line flux normalized at 19:12. While the OH lines show a decrease by a factor of 2, which is consistent with previous observations \citep[e.g.][]{1992MNRAS.259..751R,1996ApJ...464..412C,2001PASP..113..197G,2008MNRAS.386...47E}, the $\mathrm{O_2}$ lines decline by an order of magnitude. As a result, the surface brightness evolution around the 1.27 $\mu$m line significantly decreases until the middle of the night as shown in Figure \ref{fig:fluxsky}. The significant decrease of the 1.27 $\mu$m line strength is consistent with the result by satellite observation of the oxygen 1.27 $\mu$m line \citep{gaoh}. Their results also indicate that this tendency is more or less universal throughout the year, at least near the equator \citep[see Figure 2 in ][]{gaoh}. Hence, we suggest that the observation aiming to detect the oxygen absorptions should be taken after 23:00 . We note that the surface brightness in the 1.26-1.275 $\mu$m range, which avoids the strong OH lines between 1.275-1.280 (see Fig. \ref{fig:spec}) is smaller than that of 1.26-1.28 $\mu$m. Taking consideration airmass uncertainty into account, we decide to use the fiducial value of the sky background at the 1.27 $\mu$m band $\langle \Ssky \rangle_{1.27 \pm 0.01 \mu m } = 5 \times 10^{-16} \, [\mathrm{J/s/m^2/\mu m/arcsec^2}]$ for computing $\Nsky$ throughout the rest of this paper.

\subsection{Leakage vs Sky Background \label{ss:ls}}

The photo-electron from the sky background and the leakage are roughly proportional to 
$\Ssky(\lambda) \, \Delta \Omega \approx 10^{-12} \Delta \Omega \, \mathrm{[erg/s/cm^2/\mu m]}$ and $\overline{\Cont_\XY}(\lambda) \Is(\lambda) \approx \Cont \times 10^{-5.48 -0.4 m}$, respectively, for the J-band, where $m$ is the apparent magnitude. Hence the sky background dominates when $m > 16.3 + 2.5 (p-q)$, where $\Cont=10^{p}$ and $\Delta \Omega = 10^{q} \,\, \mathrm{[arcsec^2]}$. For instance, $p=-8$ and $q=-3.5$, one find $m>5.4$, that is, the sky background dominates the leakage and one can estimate
\begin{eqnarray}
(S/N)_{\biom} \approx \frac{\Nabs}{\sqrt{\NNsky + \Np}},
\label{eq:snapx}
\end{eqnarray}
for most cases. We will discuss the validity of this assumption using nearby stars catalogs in \S 3.

\subsection{Possibility of False Positive by Speckle Noise\label{ss:fps}}
 We revisit the assumption that the starlight speckles do not create an absorption-like feature in equation (\ref{fig:jcont}).  The speckle noise appears at the angular separation $\theta_{\mathrm{speckle}} (\lambda) = \lambda/x_0$ radian from the detector center due to the phase error, where $x_0$ is spatial period of a typical sinusoidal ripple of the phase of light \citep[e.g.][]{traub11}. Hence the position of speckles depends on the wavelength. We consider how the speckle moves in the range of wavelength between the oxygen band width $\Delta \lambda$. Let us describe the distance of the planet from the detector center by $\alpha \lambda/D$ in the diffraction limit unit, the speckle near the planet is made by the ripple of $x_0 \approx D/\alpha $. Then the speckle moves to the radial direction by $\Delta \lambda/x_0 = \alpha (\Delta \lambda/ \lambda) (\lambda/D) $ in the range we considered. Hence, if $\alpha < \lambda/\Delta \lambda$ the speckle does not move beyond the PSF size in the range of $\Delta \lambda$. For a 0.1 $\mu$m width around the 1.27 $\mu$m band ( $\lambda/\Delta \lambda \sim 13$), the speckle does not change for the planet at $<$ 100 mas from the center for the 30 m telescope. As shown later, the planets detectable by the ground-based telescope are generally located closer than 100 mas. In this range of planet separation, we can assume that $\Cont_\XY (\lambda)$ has the same shape as the average of contrast $\overline{ \Cont_{\XYP} } (\lambda)$. Then the wavelength dependence of the contrast should be smooth around the 1.27 $\mu$m band. The contribution of the contrast from the AO smoothly depends on $\lambda^{-2}$. Several coronagraphs have smooth wavelength dependence around 1.27 $\mu$m \citep[e.g. Apodized Pupil Lyot Coronagraph; APLC ][]{2011PASP..123...74H}, or even achromaticity such as Savart-Plate Lateral-shearing Interferometric Nuller for Exoplanets \citep[SPLINE; ][]{murakami}.  With these coronagraphs, we might be able to avoid false positives due to speckle noise. However speckle patterns also exhibit a degree of chromaticity due to the chromaticity of various components of the optical path that impact the quasi-static speckle pattern. Though we ignore the systematics due to chromaticity in this paper, we stress again the importance of the systematic due to chromaticity.

\section{Statistical Analysis of Nearby Stars \label{sec:neabycat}}

In this section, we examine the feasibility of the 1.27 $\mu$m detection from an Earth twin at the IHZ, if exists, around real nearby stars. 

\subsection{Nearby Star Catalogs}

We compile several nearby star catalogs: the all-sky catalogue of bright M Dwarfs \cite{2011AJ....142..138L} (hereafter LG11) ,  LSPM catalogue \citep{2005AJ....129.1483L}, the Hipparcos and Tycho Catalogs \citep{1997ESASP1200.....P}, NStars (\cite{2003AJ....126.2048G,2006AJ....132..161G}), and catalogs by \citet{1991adc..rept.....G}, \citet{2008MNRAS.389..585C} (C08), and \citet{2010A&A...512A..54C} (C10). All data except for LG11 \footnote{http://heasarc.gsfc.nasa.gov/W3Browse/all/mdwarfasc.html} were taken via VizieR catalogue service \citep{2000A&AS..143...23O} and are compiled so as to combine duplicate entries. 

 The stellar flux is computed from the J band magnitude if available (LG11, C08, and C10), otherwise we simply use by Planck distribution with estimated stellar temperature $\Teff$ and luminosity $L_\star$, 
\begin{eqnarray}
\Is(\lambda) &=& \frac{2 \pi h c^2}{\lambda^5} \frac{ \Rs^2}{d^2} \left[ \exp{ \left(\frac{h c}{\lambda k \Teff} \right) }- 1 \right]^{-1} \left( \frac{\lambda}{h c} \right),
\end{eqnarray}
where  $R_\star = \sqrt{L_\star/(4 \pi \sigma_\mathrm{SB} \Teff^4)}$ is the stellar radius, $\sigma_\mathrm{SB}$ is the Stefan-Boltzmann constant. 

We use $\Teff$ provided in NStars, C08, and C10, and $L_\star$ for the latter two. For the other catalogs, we estimate $\Teff$ using the color-temperature relation. We use (V-H)-$\Teff$ relation of the M dwarfs derived by \citet{2008MNRAS.389..585C} for LSPM up to $\Teff=4000$ K and (V-K)-$\Teff$ of \citet{1996A&A...313..873A} (V-K $> 0.4$) for stars of LG11 and LSPM (if not available the H magnitude),  and (B-V)-$\Teff$ by \citet{1996A&A...313..873A}  assuming the solar metalicity for the Hipparcos and Tycho Catalogs \citep{1997ESASP1200.....P} (B-V $> 0.2$). In the latter two cases, we restrict $\Teff > 3600$ K because of the approximate validated range  \citep[see Fig 1a/8a of][]{1996A&A...313..873A}. The stellar luminosity for $\Teff \ge$ 4000 K is computed from the V-band magnitude $m_V$ with the bolometric correction \citep{1996ApJ...469..355F, 2010AJ....140.1158T}, $\mathrm{BC}_V (\Teff)$. Since $\mathrm{BC}_V (\Teff)$ is not applicable for M-dwarfs,   we adopt the empirical relation of the bolometric magnitude $m_{\mathrm{bol}}$ and the H-band magnitude $m_\mathrm{H}$ derived by \citet{2008MNRAS.389..585C} to stars with $\Teff < $ 4000 K in LSPM,
\begin{eqnarray}
m_{\mathrm{bol}} = 1.94+1.06\, m_\mathrm{H}.
\end{eqnarray}
We assume a telescope on Mauna Kea and eliminate stars with declination $< -40$ degree. Finally we pick up \numstar stars within 30 pc. The numbers of M, K, G, \& F stars are 202,36,16,7 within 10 pc and 2299,879,439,262 within 30 pc, corresponding to the completeness 96, 86, 62, \& 64 \% within 10 pc and 41, 78, 63, \& 88 \% within 30 pc, respectively. 

\subsection{Planet Assumptions}

For the planet, we assume the current Earth-twin at the inner edge of the habitable zone, which has the equivalent width of the 1.27 $\mu$m band $\Woxy=2$ nm and the bond albedo $A(\lambda)$= 0.3, which is a typical value for the Earth. We assume that the planet is at maximum elongation $\beta=90$ degree. We substitute $\dsp$ by the inner habitable distance given by \citet{1993Icar..101..108K},
\begin{eqnarray}
\dsp = \dsp_\mathrm{IHZ} = \sqrt{\frac{L_\star/L_\odot}{S_\mathrm{eff}}} \mathrm{\,\,\, [AU]},
\end{eqnarray}
where $S_\mathrm{eff}=1.05$ for M and K-type stars 1.41 for G-type stars and 1.90 for F-type stars.

We use 0.95 as a fiducial value of $\Trm$, which is typical atmospheric transmission at Mauna Kea (0.93-0.96 for 1.26-1.28 $\mu$m)\footnote{http://www.gemini.edu/sciops/telescopes-and-sites/observing-condition-constraints/ir-transmission-spectra, Lord, S.D. 1992, NASA Technical Memor. 103957}.

%

\begin{table}[tbh]
\begin{center}
\caption{Fiducial parameters for observation \label{tab:fvo}}
  \begin{tabular}{cc}
   \hline\hline
   parameter  & value \\
   \hline
   \multicolumn{2}{c}{planet} \\
   \hline
   planet radius, $\Rp $ & 1 $R_\oplus$ \\  
   semi-major axis, $\dsp_\mathrm{IHZ}$ & IHZ \\
   EW of the 1.27 $\mu$m band, $\Woxy$ & 2 nm \\   
   the Bond albedo, $A(\lambda)$ & 0.3 \\
   \hline
   \multicolumn{2}{c}{observational condition} \\
   \hline
   atmospheric transmission, $\Trm $ & 0.95 \\
   airglow at 1.27 $\mu$m,  $\langle \Ssky \rangle_{1.27 \pm 0.01}$ $\mu$m & $ 5 \times 10^{-16}$ \\
   $(\mathrm{J/s/m^2/\mu m/arcsec^2})$ &  \\
   \hline
   \multicolumn{2}{c}{photon-noise limited detector} \\
   \hline
   diameter of telescope aperture, $\DA$ & 30 m \\
   Inner Working Angle  & $\lambda/\DA=8.7$ mas \\
   extracted region at the planet, $\Delta \Omega$ &  $2.4 \times 10^{-4}$  \\
   $\mathrm{(arcsec^2)}$  & \\
   throughput, $\TP$  &  0.5\\
   quantum efficiency, $\QE$  & 0.75 \\
   exposure time, $\Texp$ & 5 hours for spectroscopy \\
                          & 1 hour for detection \\
   raw contrast $C_\mathrm{raw}$ & $10^{-4}$ at 10 mas\\
                                 &    $10^{-6}$ at 100 mas  \\
   detection contrast & set by photon noise \\
                      & equation (\ref{eq:photonnoise})\\
\hline
\end{tabular}
\end{center}
\end{table}

\subsection{Instrument performance assumed}

\begin{figure}[!tbh]
\includegraphics[width= 0.9\linewidth]{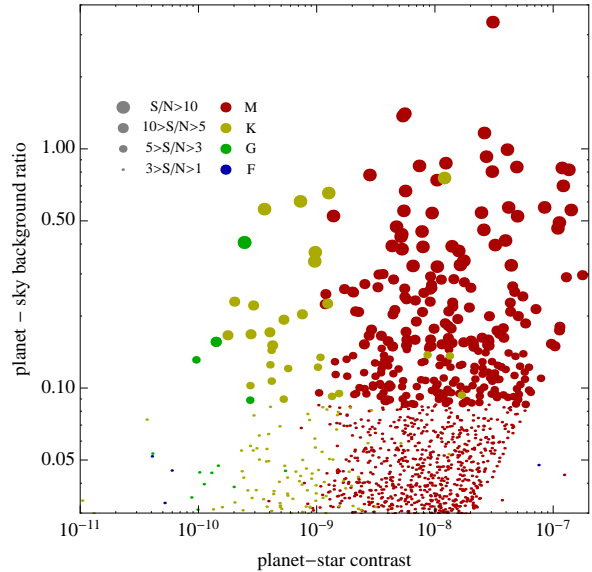}
  \caption{The ratio of the planet flux and the flux from the sky background for the J band ($\langle \Ssky \rangle$) within $\Delta \Omega = 2.4 \times 10^{-4} \mathrm{arcsec^2}$, which corresponds to the PSF circle for a 30 m telescope.  \label{fig:psr}}
\end{figure}

As an ELT, we assume a 30 m telescope at Mauna Kea, inspired from TMT. Since the planet detection strongly depends on instrument performance, we first examine the ratio of the planetary flux vs the sky background. 
Figure \ref{fig:psr} shows the ratio 
\begin{eqnarray}
c \equiv  \frac{\Ip(\lambda_c)}{\langle \Ssky(\lambda) \rangle_b } \left(  \frac{\Delta\Omega}{2.4 \times 10^{-4} \,\, \mathrm{arcsec^2}} \right)^{-1}.
\end{eqnarray}
The sky background is brighter than the planetary flux for most cases. To detect the planet, the leakage should be smaller than the planetary flux since the rms of the speckle noise is the same order as the leakage flux. The leakage does not affect the photon statistics significantly {\it if the leakage is suppressed enough to detect the planet}. Hence we discuss the feasibility of the absorption detection separately from the planet detection ignoring the photon counts from the leakage.

The planet detection is also considered assuming an idealistic instrument combined a coronagraph and post-processing. We assume the raw contrast $C_\mathrm{raw}=10^{-4}$ at 10 mas and $C_\mathrm{raw}=10^{-6}$ at 100 mas. At least in principle, the detection contrast by the postprocessing should be set by the photon noise. Since the aim of this paper is to show the detectability of oxygen for  an ideal instrument (as opposed to current instruments), we assume a photon-noise limited detection with throughput = 0.5. The inner Working Angle (IWA) adopted is  the diffraction limit.
A detailed description of the assumed instrument performance is summarized in Appendix A. For the absorption line detection, we assume a 5 hours exposure, which is reasonable from the viewpoint of the lowest sky backgrounds in one night as shown in Figure \ref{fig:fluxsky}.

\subsection{Results \label{ss:results}}

\begin{figure}[!tbh]
\includegraphics[width=\linewidth]{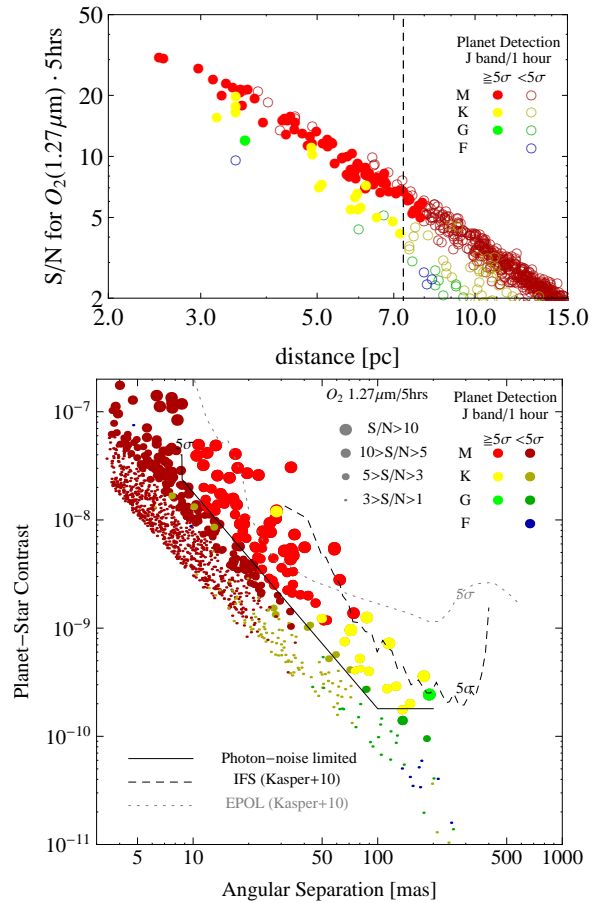}
  \caption{Detectability of the 1.27 $\mu$m band from the Earth twin at the inner habitable edge with the 30 m telescope at Mauna Kea (declination $>$ -40 degree) and 5 hour exposure. We assume that the stellar leakage is smaller than the sky background (see \S \ref{ss:ls}).  The upper panel displays the signal to noise ratio as a function of distance. Filled circles indicate the candidates with detection S/N  $> 5 \sigma$ for a photon-noise limit (1 hour observation) on TMT and with the star-planet angular separation larger than 8.7 mas. The vertical line is the approximate detection limit of the photon-noise limited detector for 1 hour exposure. The lower panel shows the signal-to-noise ratio on the angular separation - the planet-star contrast. The 5 $\sigma$ detection limits for the photon-noise limited detector,  IFS and EPOL for EPICS/E-ELT are shown by solid, dashed, and dotted lines \citep{2010SPIE.7735E..81K}. \label{fig:snbio}}
\end{figure}

Figure \ref{fig:snbio} displays the signal-to-noise ratio for the star catalogue computed by equation (\ref{eq:abssn}) assuming $\NNleak \ll \NNsky$. The linear trend of the S/N - distance relation is easily understood: since we are considering the planet at the inner edge of the habitable zone, the luminosity from the planet is constant and then the S/N is approximately proportional to $d^{-2}$. Since the J-band magnitude for the same luminosity and distance depends on the stellar temperature, the S/N is also a function of the stellar temperature, which causes the slight dependence of the S/N on the spectral type of the star. While 178 stars, which are within 10 pc, have $S/N$ above 5, we stress that this estimation assumes that the planet has already been detected.

Here we consider the planet detectability with a photon noise limited detection. We theoretically estimate the planet detectability for each hypothetical Earth-twin with the photon-noise limited instrument for a 1 hour J-band observation (see Appendix A and Eq. [\ref{eq:sndetth}]). The planets with angular separation smaller than 8.7 mas are eliminated due to the IWA. The filled circles in Figure \ref{fig:snbio} show the planets with detection S/N $ \ge 5 \sigma$ in the photon noise limit. As written in Appendix A, the detection limit depends primarily on the distance and dependence on the spectral type is not strong. This distance for 1 hour is indicated by the vertical dashed line $\sim 7$ pc.

In general, the achievable contrast on coronagraphs strongly depends on the angular separation, $\alpha$ between the planet and the star. The bottom panel of Figure \ref{fig:snbio} shows $S/N$ of the absorption band detection (indicated by the size of points) on the angular separation-planet-star contrast ($c_\mathrm{sp} \equiv \Ip/\Is$) plane. On this plane, we display the 5 $\sigma$ detection limit of the planet for proposed detectors. Solid curves indicates the detection limit of the photon-noise limited detector for a 1 hour J-band observation with TMT.

Ninety seven stars (79, 17, and 1 for M, K and G stars) for the photon-noise limited detector are above the 5 $\sigma$ detection limit and almost all these stars (76 stars) have S/N $>5$ of the oxygen band detection. Since the completeness of the M and K stars within 10 pc is high (96\% and 86\%), the number of the stars above the limit  in our sample is close to the real value.

For reference, we also plot the 5 $\sigma$ limit for currently proposed instruments. The dashed curve corresponds to the detection limit for the visible-light coronagraphic imaging polarimeter IFS (near-infrared integral-field spectrograph) for EPICS, though IFS assumes a 42 m telescope. We also show the 5 $\sigma$ detection limit of EPOL (the exoplanet polarimeter for EPICS) at the E-ELT for reference although the J-band is not used for EPOL. These detection limits for EPICS are taken from \citet{2010SPIE.7735E..81K}. 

In this paper, we do not take single/binary stars into consideration since the information about it is quite incomplete for the catalog we use. For M and K subdwarfs, the estimated values of the binary fraction are around 30 \%: $35 \pm 5$ \% \citep{1997AJ....113.2246R}, $26 \pm 3$ \% \citep{2004ASPC..318..166D}, $26 \pm 6$ \% \citep{2009AJ....137.3800J} and references therein. Since direct imaging of planets around binary stars is difficult, we should multiply our expected number of the detectable planets by 0.7. Hence we conclude that planets around $\sim 70$ single stars are above the 5 $\sigma$ detection limit and $\sim 50$ of these have S/N $>5$ for the oxygen band detection.

\subsection{Planet Detection with the Absorption Feature \label{ss:detabs}}
So far, we have discussed the feasibility of the oxygen absorption assuming the planet is already detected. However, it is valuable to consider an absorption feature of the planet candidate below 5 $\sigma$ detection since absorption feature itself increases the confidence level of the planet detection. Statistically, the combination of both the 3 $\sigma$ planet detection and the 3 $\sigma$ absorption detection provides almost the same probability as a 5 $\sigma$ detection. While this situation may be less reliable than the 5 $\sigma$ planet detect alone, the planet search with spectral anomaly such as the absorption feature will be useful for large telescopes with good photon statistics. Indeed the simultaneous differential imaging (SDI) technique has been implemented on several instruments, which utilize the 1.6 $\mu$m methane absorption feature for identification of gas giants \citep[e.g.][]{2007ApJS..173..143B}. \citet{2009PASP..121..716B} tested a non-simultaneous spectral differential imaging technique (NSDI) which utilize the oxygen 0.76 $\mu$m absorption, assuming observation of an Earth-like planet from space. Since the 1.27 $\mu$m oxygen feature is the strongest in J,H,and K bands except for water of the Earth reflection \citep{2009Natur.459..814P}, the utilization of this band to detect the Earth-like planet is reasonable. 
\section{Discussion}

In this paper, we have considered habitable planets with the oxygenic environment created by oxygenic photosynthetic organisms. Although the probability of the emergence of oxygenic photosynthetic organisms is extremely difficult to evaluate, we can compare available energy of nearby detectable stars to use the photosynthesis with that on the Earth. Since the photosynthesis process on the Earth reduces the photon energy to the specific excitation energy at the center of the reaction via the light harvesting antenna, the availability of photon energy by the photosynthetic organism is quantified by the photon flux, not by the energy flux, of the PAR, known as the Photosynthetic Photon Flux Density (PPFD). Since the excitation energy at the reaction center is $\sim 0.7$ $\mu$m, the PAR is defined in the range of 0.4-0.7 $\mu$m for the oxygenic photosynthesis on the Earth. The PPFD of the habitable planet around late-type stars is a half to a tenth of that of the Earth \citep[e.g.][]{2002Icar..157..535W,2007AsBio...7..252K}. We compute the PPFD   for the detectable planets by the photon-noise limited detector in our sample indicated in Figure \ref{fig:snbio}. While \citet{2007AsBio...7..252K} have argued the possibility to extend the energy to redder to obtain much energy on the planet around the late-type stars, which might need to the 3-photon process to obtain enough energy to make oxygen molecules from water, we conservatively take the same range as terrestrial oxygenic photosynthesis (0.4-0.7 $\mu$m). Since the central wavelength of the V-band 0.55 $\mu$m is also the center of the range of the PPFD, we extrapolate the V-band magnitude to the PAR and compute the PPFD. If the V-band magnitude is not available, we estimate the PPFD assuming the Planck distribution although the visible magnitude is systematically smaller than that expected from the black body radiation for the late-type star \citep{2007AsBio...7..252K}. As shown in Figure \ref{fig:snbio}, the PPFD of most detectable planets by the photon-noise limited detector is around 10-40 \% that of the Earth, which is not significantly low for the oxygenic photosynthesis. We adopt the PPFD of the Earth from Table A5 of \citet{2002Icar..157..535W} with the distance correction (from the IHZ to 1 AU). The PPFD we computed is the one at the top of the atmosphere, not the one at the planetary surface, which the organism can use. Cloud coverage and optical depth are also important to determine the PPFD at the planetary surface. These might depend on many factors, such as the spin rotation period, water content, surface temperature and so on. Observationally, if the planet is tidally unlocked, the photometric variation can be used to estimate the cloud coverage \citep{2001Natur.412..885F,2011ApJ...738..184F} or even mapping the cloud distribution \citep{2011ApJ...739L..62K, 2012arXiv1204.3504F}. For tidally locked planet, the cloud coverage for the illuminated side might be larger than that of the Earth \citep[e.g.][]{2003AsBio...3..415J,2011ApJ...743...41K}. Thus, availability of the photosynthetic light on planets around the late-type star is still an open question for both aspects of cloud formation and methodology to estimate for tidally locked planets.

\begin{figure}[!tbh]
\includegraphics[width= 0.9\linewidth]{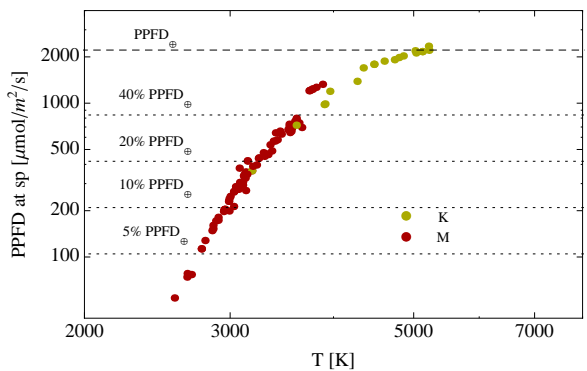}
  \caption{Photosynthetic Photon Flux Density (PPFD) of the detectable planets at the inner habitable zone. The range of the Photosynthetically Active Radiation (PAR) is 0.4-0.7 $\mu$m. The PPFD for filled and symbols are computed from the V-band magnitude while opened ones are obtained as black body radiation due to lack of the V-band magnitude. Circles and rectangles represent the planets above 5 $\sigma$ detection by the photon-noise limited detector. The horizontal line indicates the PPFD on the Earth, PPFD$_\oplus$ \citep{2002Icar..157..535W}. The three dotted lines correspond to 40, 20, 10, and 5 \% of the Earth PPFD level from top to bottom.  \label{fig:par}}
\end{figure}

We did not consider detectability of other important biomarkers including $\mathrm{H_2 O, CH_4, CO_2}$ in this paper.  The detection of oxygen with reductant such as methane might support the evidence of the existence of exo-life. In this case, methane does not have to be biotic molecules. It proves that there exists some mechanism which continuously produces oxygen, since oxygen is consumed in the oxidation of methane. While methane on the current Earth is more challenging to detect, \citep{2005AsBio...5..706S} showed that UV photochemical process significantly enhances the absorption of methane for habitable planets around late-type stars (Figure 8 in their paper). As shown in \citep{2005AsBio...5..706S}, the detectability of methane is complex because the methane is minor component in the current Earth (but oxygen is not the case). Since the issue that abiotic production of oxygen is still under debate, we do not consider the methane detection in this paper. The practical problem is that we can no longer adopt the simple estimate of $S/N$ since these lines are not in the middle of bands and, have wider wavelength range. We will consider detectability of these biomarkers by generalizing the method in the next paper.

\section{Conclusion}

In this paper, the detectability of oxygen absorption, as a biomarker from habitable planets, by ground-based telescopes has been examined. Analyzing real data of the night sky, we showed that the telluric oxygen emission as the strong sky background is damped by an order of magnitude during a night. With nearby star catalogs, which is almost complete for M and K stars within 10 pc, we found that the oxygen 1.27 $\mu$m absorption band is detectable with future photon noise limited instruments and 30-40 m telescopes for planets around dozens to hundreds late-type stars assuming an Earth-twin at the inner habitable zone around the stars. Photosynthetic photon flux density of these planets is not significantly low, 10\%-40\% of that on the Earth. We conclude that extremely large telescopes in the near future will enable us to search for the signature of oxygenic photosynthesis on the habitable planets around the late-type stars. This paper demonstrates the importance of deploying small IWA ($\sim \lambda/D$) efficient coronagraph + ExAO on ELTs, and clearly shows that doing so will enable study of potentially habitable planets.

We are deeply grateful to Enric Pall{\'e} for allowing us to use the Earthshine data. We are greatly thankful to Fumihide Iwamuro for discussion about telluric oxygen lines and Kintake Sonoike and Tae-Soo Pyo for helpful and insightful discussion. We also an anonymous referee for a lot of constructive comments, which significantly improved the manuscript. HK is supported by a JSPS (Japan Society for Promotion of Science) Grant-in-Aid for science fellows. This work is also supported by Grant-in-Aid for Scientific research from JSPS and from the Japanese Ministry of Education, Culture, Sports, Science and Technology (Nos. 22$\cdot$5467, 23$\cdot$6070, 23740139, 24103501, 22000005) and National Science Council of Taiwan (No. 100-2112-M-001-007-MY3).

\appendix

\section{Theoretical limit of the planet detection for the photon noise limited case}
Here we examine the theoretical limit of the planet detectability for the photon noise limited case. We assume that the coronagraph first suppresses the photon of the stellar leakage to $N = C_\mathrm{raw} \Ns$ at the planet location, where $C_\mathrm{raw}$ is the raw contrast and $\Ns$ is the photo-electron of the main star. 

For the photon-noise limited case, dynamic range is limited by the photo-electron counts of the planet through the coronagraph
\begin{eqnarray}
 dyn = \sqrt{\eta N},
 \label{eq:photonnoise}
\end{eqnarray}
where $\eta$ is a detector-dependent factor. We use $\eta=1/2$, which is the same value estimated for the Pupil Remapping Imager \citep{2006MNRAS.373..747P}. Then, the signal-to-noise ratio is expressed as 
\begin{eqnarray}
(S/N)_\mathrm{det} = \frac{\Np}{\sqrt{\eta C_\mathrm{raw} \Ns}} = c_\mathrm{sp} \sqrt{\frac{\Ns}{\eta C_\mathrm{raw} }}.
\label{eq:sndetth}
\end{eqnarray}
 The raw contrast is primary determined by the ExAO. 
\citet{2010aoel.confE3007K} performed the simulation of ExAO for EPICS and showed that ExAO can achieve the contrast $\sim 10^{-4}$ and $\sim 10^{-6}$ at the angular separation 10 mas and 100 mas, respectively (see Fig. 4 of their paper). Hence we assume the raw contrast within IWA as a function of the angular separation as
\begin{eqnarray}
C_\mathrm{raw} = \mathrm{max} \left\{ k \left( \frac{\alpha}{\mathrm{mas}} \right)^{-2}, 10^{-6} \right\},
\end{eqnarray}
where we adopt $k=10^{-2}$.  The condition of the 5 $\sigma$ detection  $(S/N)_\mathrm{det} \ge 5$ for the Earth-twin at the IHZ can be rewritten as
\begin{eqnarray}
\sqrt{\frac{2 N_\star}{C_\mathrm{raw}}} \left( \frac{R_\oplus}{a_\oplus}\right)^2 \left( \frac{L_\star}{L_\odot}\right)^{-1} S_\mathrm{eff} \phi (\beta) A (\lambda) \ge 5.
\label{eq:qqq}
\end{eqnarray}
Writing 
\begin{eqnarray}
N_\star \approx N_{\odot,10} \left( \frac{L_\star}{L_\odot}\right) \left( \frac{d}{10 \mathrm{\, pc}}\right)^{-2} f(\Teff)
\end{eqnarray} 
where  $N_{\odot,10}$ is the photon count of the Sun observed at the distance of 10 pc with a considered instrument and $f(\Teff)$ is a correction factor due to the stellar temperature dependence between photon counts and luminosity. Equation (\ref{eq:qqq}) reduces
\begin{eqnarray}
d \le 2 \times 10^{-3} [N_{\odot,10} f(\Teff)]^{1/4} \mathrm{\, pc} \sim 7.3 \mathrm{\, pc},
\end{eqnarray} 
where the last value is derived on the assumption of 5 hours exposure with the instruments in Table \ref{tab:fvo} and $f(\Teff)=1.9$ for $\Teff=3000$ K (we assume the black body).


\end{document}